\documentclass[%
 reprint,
 amsmath,amssymb,
 aps,
]{revtex4-1}

\usepackage{graphicx}
\usepackage{dcolumn}
\usepackage{bm}
\usepackage{amsmath}
\usepackage{amssymb}
\usepackage{physics}
\usepackage{booktabs}

\begin{document}

\preprint{APS/123-QED}

\title{An Upper Bound on the Strongly Forbidden $6S_{1/2} \leftrightarrow 5D_{3/2}$ Magnetic Dipole Transition Moment in {Ba}$^{+}$ }

\author{Spencer R. Williams }
\affiliation{%
 Department of Physics, University of Washington, Seattle, Washington, 98195, USA\\
 }%
\author{Anupriya Jayakumar}%
\affiliation{%
 Department of Physics, University of Washington, Seattle, Washington, 98195, USA\\
}%

\author{Matthew R. Hoffman}%
\affiliation{%
 Department of Physics, University of Washington, Seattle, Washington, 98195, USA\\
}%

\author{Boris B. Blinov}%
 \email{blinov@uw.edu}
\affiliation{%
 Department of Physics, University of Washington, Seattle, Washington, 98195, USA\\
}%

\author{E.N. Fortson}%
\affiliation{%
 Department of Physics, University of Washington, Seattle, Washington, 98195, USA\\
}%

\date{\today}

\begin{abstract}
	    \noindent We report the results from our first-generation experiment to measure the magnetic-dipole transition moment (M1) between the $6S_{1/2}$ and $5D_{3/2}$ manifolds in Ba$^{+}$.  
	    Knowledge of M1 is crucial for the proposed parity-nonconservation experiment in the ion \cite{Fortson93}, where M1 will be a leading source of systematic error. To date, no measurement of M1 has been made in Ba$^{+}$, and moreover, the sensitivity of the moment to electron-electron correlations has confounded accurate theoretical predictions.  A precise measurement may help to resolve the theoretical discrepancies while providing essential information for planning a future PNC measurement in Ba$^{+}$.  We demonstrate our technique for measuring M1 - including a method for calibrating for stress-induced birefringence introduced by the scientific apparatus - and place an upper bound of $\mathrm{M1} < 93 \pm 39 \times 10^{-5} \mu_{B}$.    
	    
\begin{description}
  \item[PACS numbers] 31.30.jc, 31.30.jg, 32.70.Cs, 32.70.Fw	
\end{description}
\end{abstract}

\pacs{Valid PACS appear here}

\maketitle

\section{Introduction}
Atomic parity nonconservation (APNC) measurements continue to be of interest for the breadth of physics they may illuminate \cite{Marciano90, Fortson90, Fortson92, Flambaum84}. 
Presently, several experiments are underway to search for parity violating effects in atoms and molecules \cite{Gwinner2007, DeMille08, Tsigutkin10, Williams13, Roberts14, Choi16}.
One path being pursued for next-generation APNC studies is to investigate the effect in single trapped atomic ions, following the approach proposed in \cite{Fortson93}, for which Ba$^{+}$ is 
exemplar.  
In that system, the largest observable consequence of APNC is the emergence of a small electric-dipole transition moment (E1$_{APNC}$) between the ion's 6S$_{1/2}$ ground states
and its low-laying metastable 5D$_{3/2}$ states.     
In addition to E1$_{APNC}$, there is also an electric-quadrupole transition moment (E2) and magnetic-dipole transition moment (M1) connecting those states.
To measure E1$_{APNC}$, the size of these other, much larger, transition moments must be known.
Some work has been undertaken that placed E2 at 12.7 ($a_{0}/\lambda$) $e a_{0}$ \cite{Sahoo06, Yu97}, where $e$ is the electron charge and $a_{0}$ is the Bohr radius,
however, to the best of our knowledge, no measurement of M1 has been done and there is significant disagreement among the numerical studies \cite{Sahoo06, Gossel13, Safronova14Priv}.  
In this paper we present the results of our experimental study of M1 and place an upper bound on the moment. \\
\section{Background}
\indent The principal challenge in measuring the 6S$_{1/2} \leftrightarrow $ 5D$_{3/2}$ M1 transition matrix element in Ba$^{+}$ is that its effect must be isolated from the much larger E2 coupling.  
These two transition moments explicitly are,
\begin{subequations}\label{eq:M1E2}
  \begin{equation}\label{eq:E2}
    \mathrm{E2} = \big\langle 6S_{1/2} \big|\big| \widehat{\mathrm{E2}} \big|\big| 5D_{3/2} \big\rangle
  \end{equation}
  \begin{equation}\label{eq:M1}
    \mathrm{M1} = \big\langle 6S_{1/2} \big|\big| \widehat{\mathrm{M1}} \big|\big| 5D_{3/2} \big\rangle
  \end{equation}
\end{subequations}
where the operators $\widehat{\mathrm{E2}}$ and $\widehat{\mathrm{M1}}$ are defined as,
\begin{subequations}\label{eq:M1E2op}
	\begin{equation}\label{eq:E2op}
		\widehat{\mathrm{E2}} = -\frac{1}{6}\widehat{Q}_{i,j} \frac{\partial E_{i}}{\partial x_{j}}
	\end{equation}
	\begin{equation}\label{eq:M1op}
		\widehat{\mathrm{M1}} = -\widehat{\vec{M}}\cdot\vec{B}
	\end{equation}
\end{subequations}
Where $\widehat{Q}_{i,j}$ and $\widehat{\vec{M}}$ are the (electric) quadrupole and (magnetic) dipole operators, respectively, and $\vec{E}$ and $\vec{B}$ are the applied electric and magnetic fields.  The matrix element E2 is known, to the 1 \% level, to be 12.7 ($a_{0}/\lambda$) $e a_{0}$ \cite{Sahoo06, Yu97}.  
However, to the best of our knowledge, only three calculations of M1 have been performed which, in terms of the Bohr magneton $\mu_{B}$, have given; 
\begin{equation}
  \mathrm{M1} =\begin{cases}
	    \,\,80 \times 10^{-5} \,\,\mu_{B} \,\,\,\, \text{\cite{Sahoo06}}\\
	    \,\,22 \times 10^{-5} \,\,\mu_{B} \,\,\,\, \text{\cite{Gossel13}}\\
	    \,\,17 \times 10^{-5} \,\,\mu_{B}  \,\,\,\, \text{\cite{Safronova14Priv}}
	  \end{cases}
\end{equation}
The large range of values obtained in those studies has been attributed to those author's estimations of electron-electron correlation effects in the atom \cite{Safronova14Priv}.  That discrepancy withstanding, there is agreement on its order of magnitude, 
which places M1/E2 at $\sim$ $10^{-3}$. \\
\indent For any of the $6S_{1/2} \leftrightarrow 5D_{3/2}$ transitions in Ba$^{+}$, which have their resonances nominally at 2.051 $\mu$m, the total Rabi frequency ignoring APNC, $\Omega$, is,
\begin{equation}
  \mathrm{\Omega = |\Omega_{E2} + \Omega_{\text{M1}}|}
\end{equation}
We seek M1 within the $\Delta$ m = 0 transition between the 6S$_{1/2}$(m = -1/2) and 5D$_{3/2}$(m = -1/2) states, for which, in terms of the reduced transition matrix elements of Eq. (\ref{eq:M1E2}), 
the E2 and M1 contributions are,
\begin{subequations}\label{eq:Rabiform}
  \begin{equation}\label{eq:RabiformB}
	  \mathrm{\Omega_{E2}^{(0)}} = \frac{i k}{4 \hbar}\sqrt{\frac{1}{10}} \mathrm{E2}\,\sin(2\theta)\,\mathrm{E}_{||}
  \end{equation}
  \begin{equation}\label{eq:RabiformC}
	  \mathrm{\Omega_{\text{M1}}^{(0)}} =  -\frac{1}{\hbar}\sqrt{\frac{1}{6}}\mathrm{M1}\,\sin(\theta)\,\mathrm{B}_{||},
  \end{equation}
\end{subequations}
where the superscript identifies the change in the magnetic quantum number for the transition, $i$ is the complex unit, and $k$ is the driving field's wave-vector.
The angle $\theta$ is that between the 2.051 $\mu$m beam and the ion's quantization axis. The $||$ and $\perp$ delimiters on the driving field components describe their orientation relative to the plane spanned by those vectors, with $||$ being the in-plane component and $\perp$ the normal component.  It follows that by driving the $\Delta m$ = 0 transition with a perfectly linear polarized field, the electric-quadrupole coupling can be turned off by rotating the 2.051 $\mu$m beam's polarization to an orientation where its electric field is perpendicular to the quantization axis. \\
\indent The alignment angle $\theta$ cannot be distinguished well within the $\Delta m = 0$ transition since, independent of the ratio of M1 to E2, it also affects the relative size of the two Rabi frequency contributions. Its effect, however, is clearly discernible within the $\Delta m$ = $\pm 2$ transitions.  For those there is only electric field coupling, but to both the $E_{||}$ and $E_{\perp}$ components of the laser's field.  Generically, the Rabi frequency for $\Delta m$ = $\pm$ 2 is: 
\begin{equation}\label{Eq:Omega2}
  \Omega^{(\pm 2)} = \frac{k}{2\sqrt{30}\hbar}\,E2\, \Big| \frac{1}{2} \sin(2\theta)E_{||} \mp i\sin(\theta) E_{\perp} \Big|
\end{equation}
Observe also that Eq. (\ref{eq:Rabiform}) and Eq. (\ref{Eq:Omega2}) hold true regardless of the 2.051 $\mu$m polarization state.  Because the M1 coupling is so small relative to the E2 coupling, distortions to the driving field polarization state are a leading systematic concern.  In practice we have found that the small amounts of ellipticity induced from birefringence of the optical viewport leading into the trapping apparatus are significant and must be accounted for in the analysis.   

\section{Apparatus Summary}
Single ions were trapped inside a linear Paul trap, described in \cite{Williams13}, with $\sim$440 kHz axial and $\sim$1 MHz radial trapping frequencies. Micromotion was compensated for by means of a disk electrode beneath the trap and a rod electrode parallel to the trap axis.
A small hole bored through the center of the disk electrode allowed neutral barium flux to reach the trap from an oven below.  The trap was loaded by a two step photo-ionization process \cite{Steele07}.  
The neutral atoms were first excited along their narrow inter-combination line at 791 nm connecting the ground states to the 6S6P $^{3}$P$_{0}$ excited states.   This transition is sufficiently narrow to allow for isotope selective loading.  
From the excited level the atoms were ionized with a 337 nm photon from a N$_{2}$ laser.  \\
\begin{figure}
  \centering
  \includegraphics[scale=0.33]{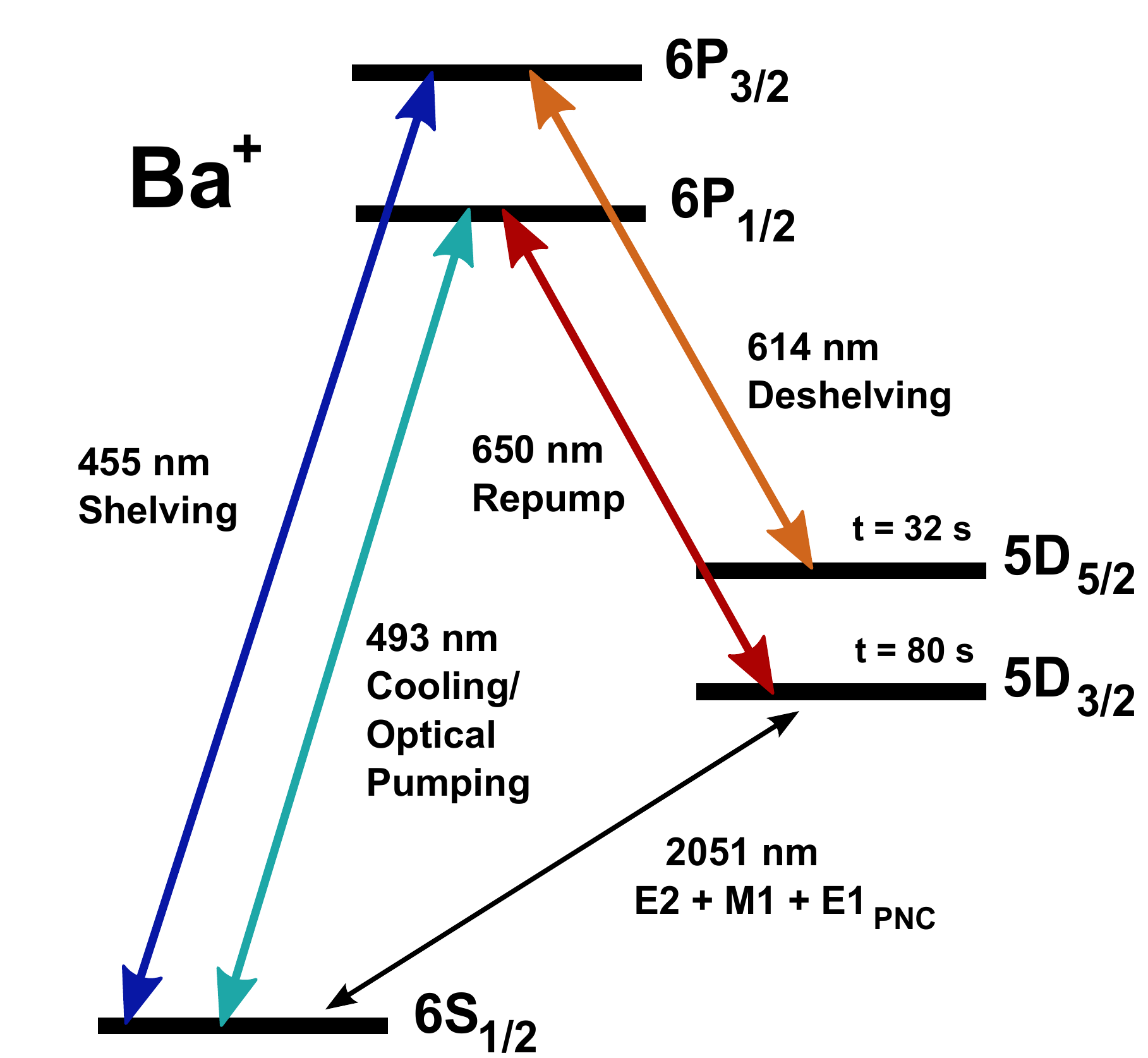}
  \caption{A diagram of the valence structure of Ba$^{+}$ including the various transitions used in the course of this experiment.}\label{Fig.BaIonDiagram}
\end{figure}
\indent The single trapped ions were cooled to $\sim$ 2 mK by Doppler cooling along the ion's 6S$_{1/2} \leftrightarrow$ 6P$_{1/2}$ transition at 493 nm.  The ion's $\Lambda$-structure, represented in Fig. (\ref{Fig.BaIonDiagram}) along with the transitions relevant to this study, necessitates a second ``repump`` beam at 650 nm.
The primary cooling beam was produced by frequency-doubling an external cavity diode laser (ECDL) operating at 986 nm.  A small percentage of the main cooling beam
was picked off and circularly polarized for optical pumping.  The 650 nm beam was produced by a commercial 650 nm ECDL.  Both beams were stabilized against highly isolated optical resonators. \\
\indent State read-out along the 6S$_{1/2} \leftrightarrow$ 5D$_{3/2}$ transitions was accomplished using an electron shelving technique, described generically in \cite{Nagourney86} and specifically for this experiment in \cite{Williams13}.  
Shelving was accomplished by means of a home built 455 nm ECDL.  Because of the strength of the transition it was convenient to operate the 455 nm laser multi-mode, which enabled us to 
drive the shelving transition without active wavelength stabilization.  Using this scheme we were able to reach the shelved state in under 5 ms, which was limited by the speed of that system's mechanical shutter.
To remove the ion from the shelved state we used a frequency-doubled 1228 nm beam producing a beam at 614 nm.  The free running stability of that laser was sufficient that it, too, did not require stabilization against an optical resonator. \\
\indent The 6S$_{1/2} \leftrightarrow$ 5D$_{3/2}$ transitions, which contain the M1 moment of interest, were excited using a diode pumped solid-state (TmHo:YLF) laser operated at 2.051 $\mu$m \cite{Kleczewski11}.
The laser was stabilized against an ultra-low expansion optical resonator that was measured to have a finesse greater than 350,000 \cite{Notcutt05, KleczewskiThesis}.  Because of the availability of optical coatings at the time of its construction, to achieve such a high finesse cavity it was necessary to stabilize it against the beam's second harmonic at 1.025 $\mu$m. 
Of the 40 mW of 2.051$\mu$m light available, only 3 mW were delivered to the ion for spectroscopy, while the rest was used for stabilization.  At the time of writing our best bound on the laser's bandwidth -
which was set in the course of this work - was lower than 70 Hz. \\
\begin{figure}
  \centering
  \includegraphics[scale=0.06]{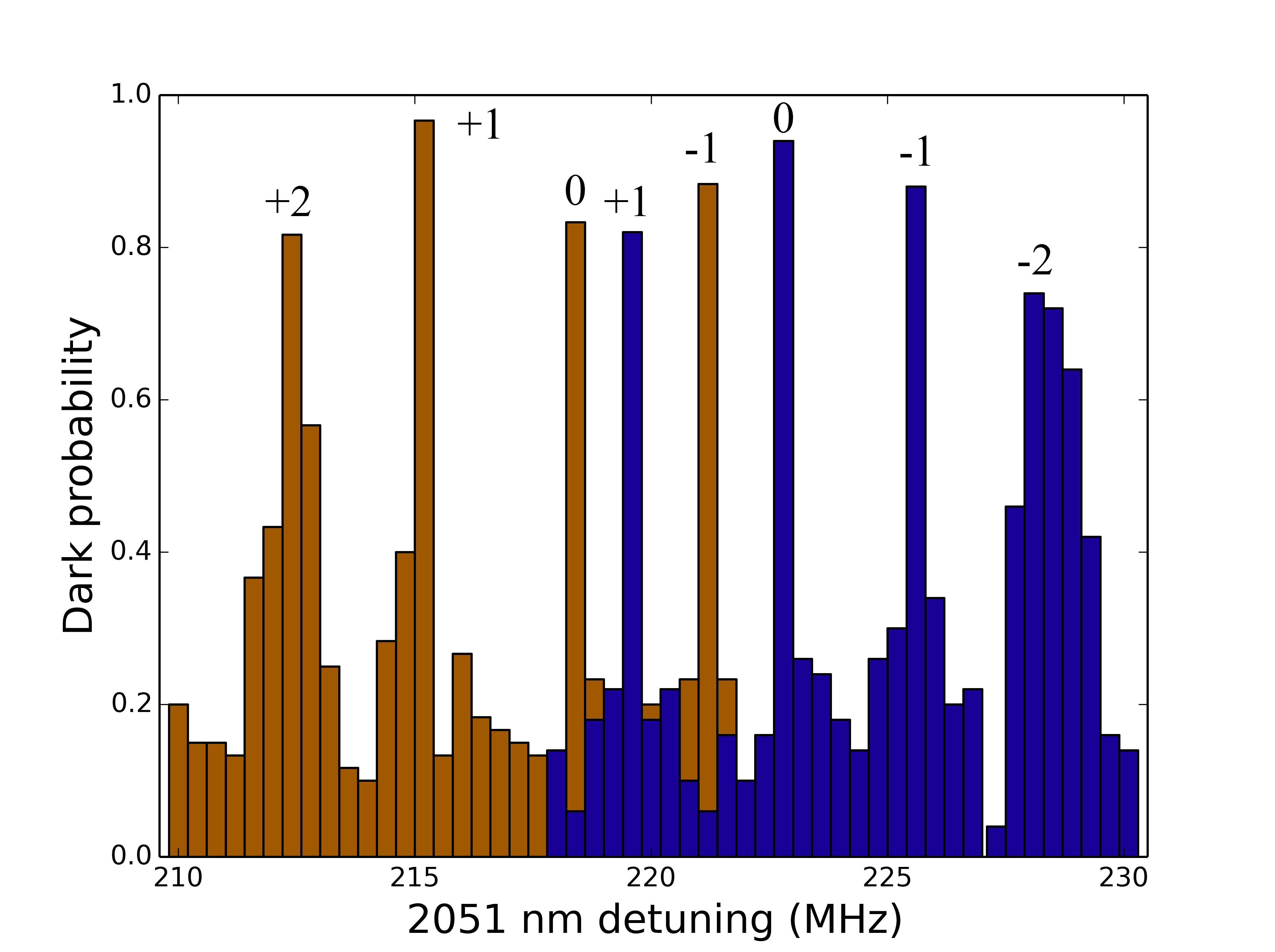}
  \caption{(Color online) Data from two adiabatic rapid passage scans that identify all eight 2.051 $\mu$m transitions through the probability of finding the ion in the dark state plotted against our frequency offset to the 2.051$\mu$m beam's nominal frequency.  Each bin is 400 kHz wide, corresponding to the width of the ARP sweep.  For the orange scan we optically pumped to the 6S($m_{j} = -1/2$) state and for the blue to the 6S($m_{j} = +1/2$) state. From left to right the tall orange bins correspond to $\Delta m = $ +2, +1, 0, -1, the blue to $\Delta m =$ +1, 0, -1, -2.}\label{Fig.RAP2um}
\end{figure}
The 2.051 $\mu$m transitions were separated by about 3 MHz by the Zeeman effect, and were typically power broadened to between tens of kilohertz to tens of hertz, depending on the field alignment.  
To roughly locate the transitions we used adiabatic rapid passage (ARP) \cite{Grischkowsky76, Noel12}.  Examples of two ARP scans, corresponding to starting with the atom prepared in either of its $6S_{1/2}$ ground states, are shown in Fig. (\ref{Fig.RAP2um}).  Within those scans all eight 2.051 $\mu$m transitions are evident, and are labeled by their corresponding change in the magnetic quantum number.
\section{Data and analysis}
\begin{figure}
  \centering
  \includegraphics[scale=0.35]{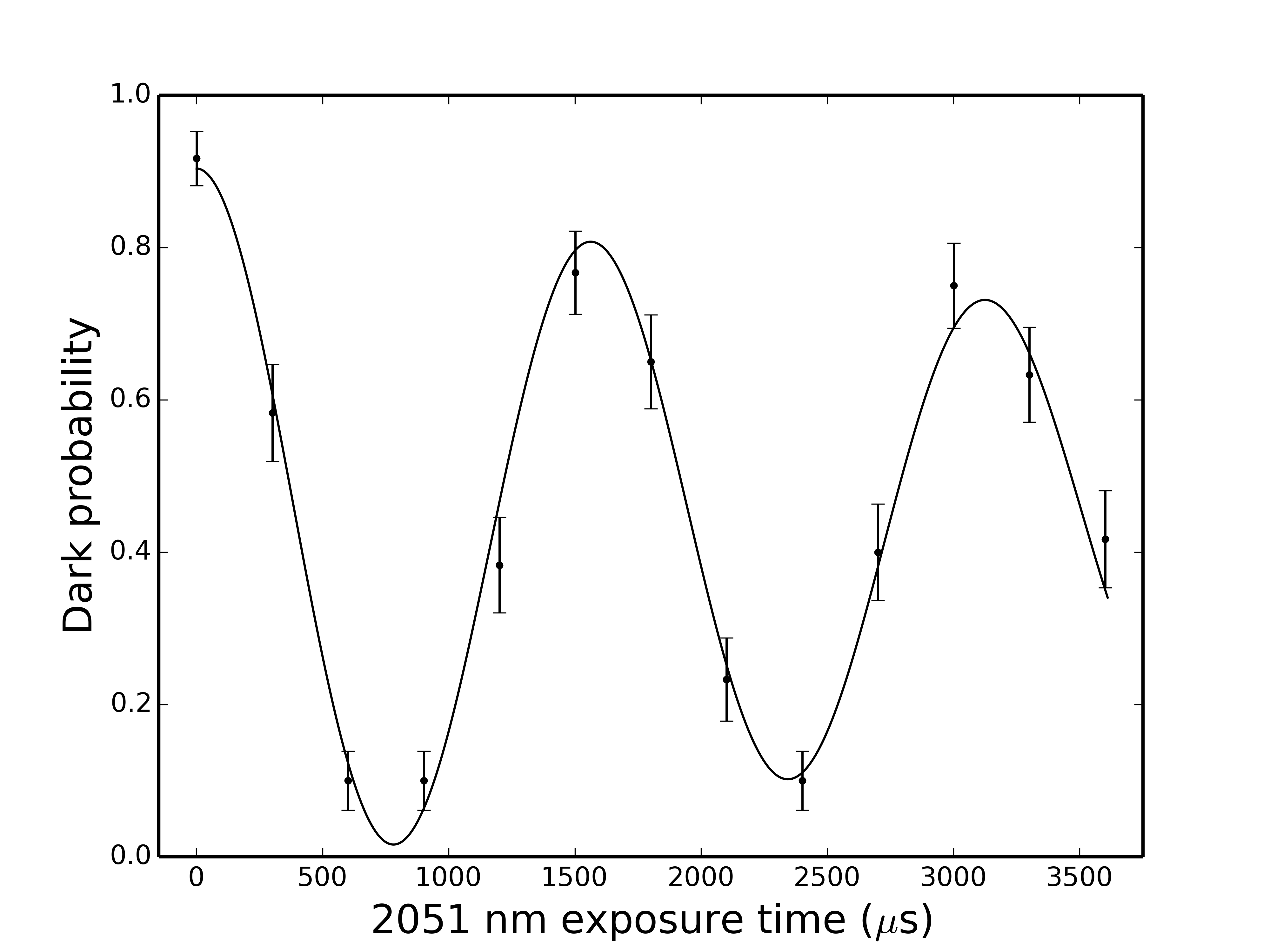}
  \caption{An example of a Rabi oscillation taken along the 2.051 $\mu$m $\Delta$m=0 transition.  The probability of finding the ion in the dark state is shown against the 2.051 $\mu$m exposure time.  For each exposure time the measurement was repeated one hundred times and the data were fit to a theoretical model accounting for the efficiency of shelving \cite{Williams13}.}\label{Fig.SampleRabiFlop}
\end{figure}

To obtain all of the necessary pieces needed to get M1, three measurements were undertaken; first of $\Omega^{(0)}$, then $\Omega^{(+2)}$ alone, then $\Omega^{(\pm 2)}$ measured simultaneously.  All Rabi frequency measurements were performed using the electron-shelving technique described in \cite{Williams13};
a sample Rabi oscillation is displayed in Fig. (\ref{Fig.SampleRabiFlop}).  The first two Rabi oscillation data sets were used to calculate M1 without correcting for 2.051 $\mu$m ellipticity, and the $\Omega^{\pm 2}$ data were used to extract the viewport birefringence parameters to account for ellipticity in the beam.  In all cases the Rabi frequency measurements were taken at 10$^{\circ}$ intervals of half-wave plate rotation, except near the minimum of $\Omega^{(0)}$, where more data was needed.  The half-wave plate position was set manually using a Thorlabs PRM1 high-precision rotation mount, which had an angular resolution of five arc-minutes read off a vernier scale.  All Rabi frequency measurements were recorded with respect to the reading of the half-wave plate's rotation mount which were assigned a $\pm$ 10 arc-minute (0.16$^{\circ}$) error bar to over-estimate error from vernier acuity.  The rotation was always carried out from low angle to high angle to avoid inconsistency from backlash.	We did not know \emph{a priori} how the half-wave plate was oriented in its rotation mount (with respect to the plane spanned by the 2.051 $\mu$m $\vec{k}$ vector and the ion's quantization axis), so in all cases the data were fit with an additional phase offset $\phi_{0}$ to account for this initial alignment. \\ 
\begin{figure}
  \centering
  \includegraphics[scale=0.35]{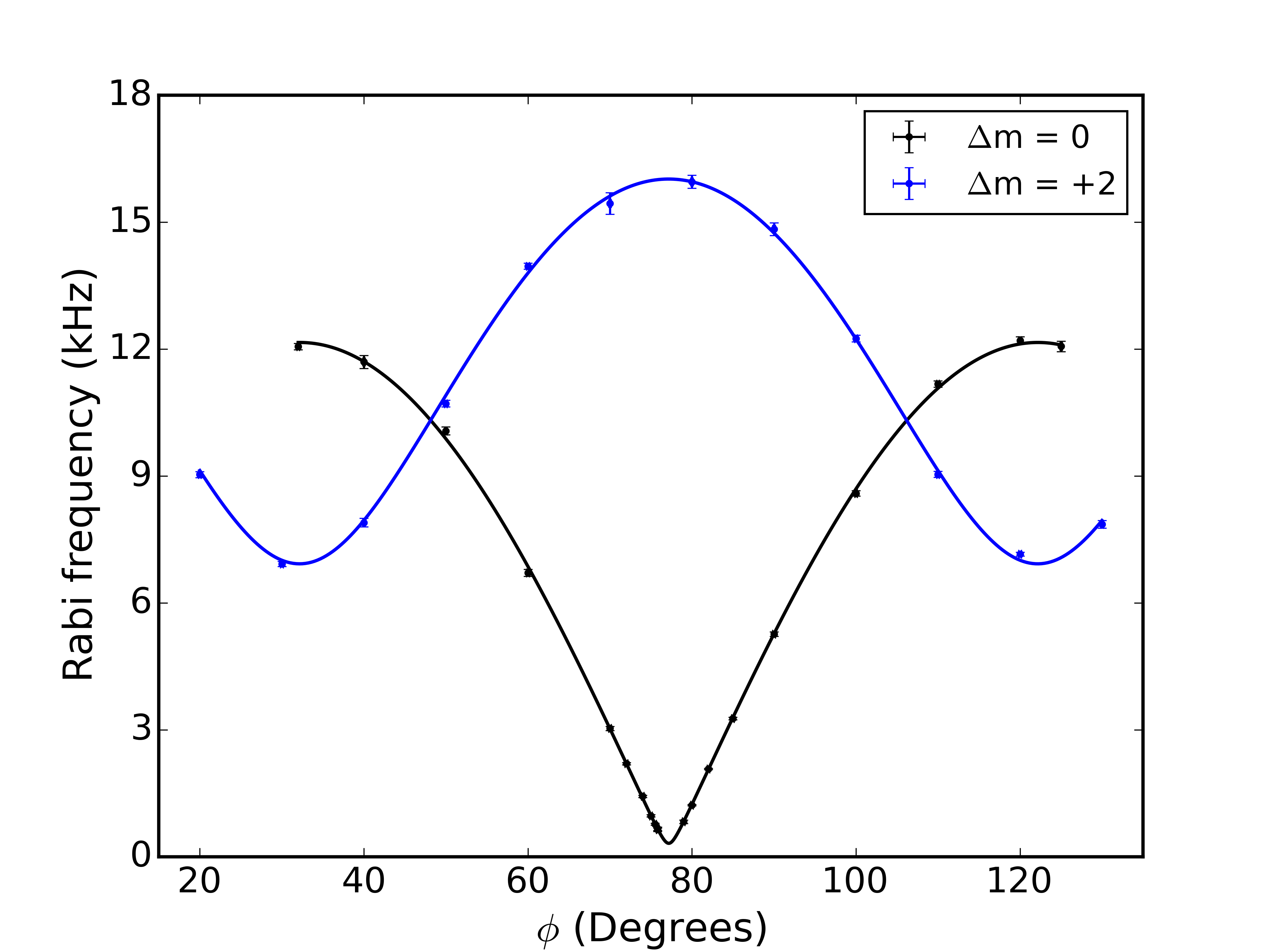}
  \caption{(Color online) Plots of the $\Omega^{(0)}$ and $\Omega^{(+2)}$ data sets along with their fits, which are described in the text.  The curve in blue is the $\Delta$m = +2 transition and that in black is the $\Delta$m = 0 transition.}
  \label{Fig:PolRotApr14Data}
\end{figure}
\indent  The data for $\Omega^{(+2)}(\phi)$ and $\Omega^{(0)}(\phi)$ are plotted together in Fig. (\ref{Fig:PolRotApr14Data}).  
For the small amounts of ellipticity we encountered, the principle effect of the viewport in the $\Omega^{(0)}(\phi)$ data is to raise the minimum value observed, $\Omega_{min}^{(0)}$, without significantly skewing its position.  The consequence of these observations is that, to a good approximation,  $\Omega^{(0)}$ is, 
\begin{equation}
  \Omega^{(0)} \approx \sqrt{ |\Omega^{(0)}_{max}|^{2}\sin^{2}(2\phi-2\phi_{0}) + |\Omega^{(0)}_{min}|^{2}\cos^{2}(2\phi-2\phi_{0}) }
\end{equation}
The parameter $\Omega_{min}^{(0)}$ includes the M1 coupling plus some E2 coupling due to unintended elliptical polarization that can be subtracted out if the viewport parameters are known.  In the data we were unable to properly resolve $\Omega_{min}^{(0)}$ because of the system's decoherence, and for that reason what we report here is an upper limit on M1.  To get that limit (first, without the ellipticity correction) we need the alignment angle between the 2.051 $\mu$m laser and the ion's quantization axis $\theta$ which we get from the stand-alone measurement of $\Omega^{(+2)}$. \\
\indent The angle $\theta$ is a fit parameter to that data and determines the amplitude of the oscillation observed $\Omega^{(+2)}(\phi)$.  The effect of ellipticity in the beam's polarization is to skew the data both vertically and horizontally which shifts $\theta$'s apparent value, however that effect contributes minimally to our final M1 value.  And so to obtain $\theta$ we approximate the laser fields as linearly polarized, which leads to,
\begin{equation}
	\begin{split}
		\Omega^{(+2)} &\approx C_{E2}\,\sin(\theta) \\
		&\sqrt{ \cos^{2}(\theta)\sin^{2}(2\phi-2\phi_{0}) + \cos^{2}(2\phi-2\phi_{0}), }
	\end{split}
\end{equation}
where $C_{E2}$ is $\frac{k|E|E2}{2\sqrt{30}\hbar}$.  
Once we have the viewport parameters we can correct $\theta$ to get closer to the true value, where we will find about a 1$^{\circ}$ shift.  That withstanding, M1 is not particularly sensitive to $\theta$, at least for our chosen geometry, so this correction is not critical for the experiment's interpretation.  An important indicator that our approximations are valid is that the data sets remain well-aligned, indicated at the fitted values of $\phi_{0}$ agreement to a few hundredths of a degree for the two transitions. \\
\begin{table}
  \caption{Fit parameters for the $\Omega^{(0)}(\phi)$ and $\Omega^{(+2)}(\phi)$ measurements} 
  \label{Table:Rabis0P2FitParameters}
  \centering 
  \begin{tabular}{c c c c c} 
	  \hline\hline\vspace{2 mm} 
	  $\Omega^{(0)}$ & $\Omega^{(0)}_{max}/2\pi$ (Hz) & $\Omega^{(0)}_{min}/2\pi$ (Hz) & $\phi_{0}$ (Degrees) \\
	  \hline
	  & $12156 \pm 35$ & $319 \pm 34$ & $77.18^{\circ} \pm 0.02^{\circ}$ \\
	  & & & \\
	  \hline\hline 
	  \vspace{2 mm}
	  $\Omega^{(+2)}$ & $C_{E2}$ (Hz) & $\theta$ (Degrees) & $\phi_{0}$ (Degrees) \\
	  \hline 
	  & $17764 \pm 81$ & $64.4 \pm .3$ & $77.16^{\circ} \pm 0.19^{\circ}$ \\ 
	  \hline\hline 
  \end{tabular} 
\end{table}	  
Solving for M1 from Eq. (\ref{eq:Rabiform}), and inserting the fit parameters in Table (\ref{Table:Rabis0P2FitParameters}) we find that 
\begin{equation}
  \begin{split}
	  \mathrm{M1} &< \sqrt{\frac{3}{5}} \times \frac{c\,k\,E2}{2} \times \frac{\Omega^{(0)}_{min}}{\Omega^{(0)}_{max}} \times \cos(\theta) \\ 
	  &=  246 \pm 26 \times 10^{-5} \mu_{B} \quad \quad (\text{No ellipticity correction})
  \end{split}
\end{equation}   
\indent To reach this value we have ignored any effects from ellipticity in the laser fields, which was nearly all introduced by the ultra-high vacuum viewport.  In the following section we will account for ellipticity in the beam's polarization by inferring the effective retardance, $\Gamma$, and orientation of the optical axis, $\alpha$, of that viewport with an \emph{in situ} measurement. \\
\section{Calibration of Viewport Birefringence}				%
\begin{figure}
  \centering
  \includegraphics[scale=0.4]{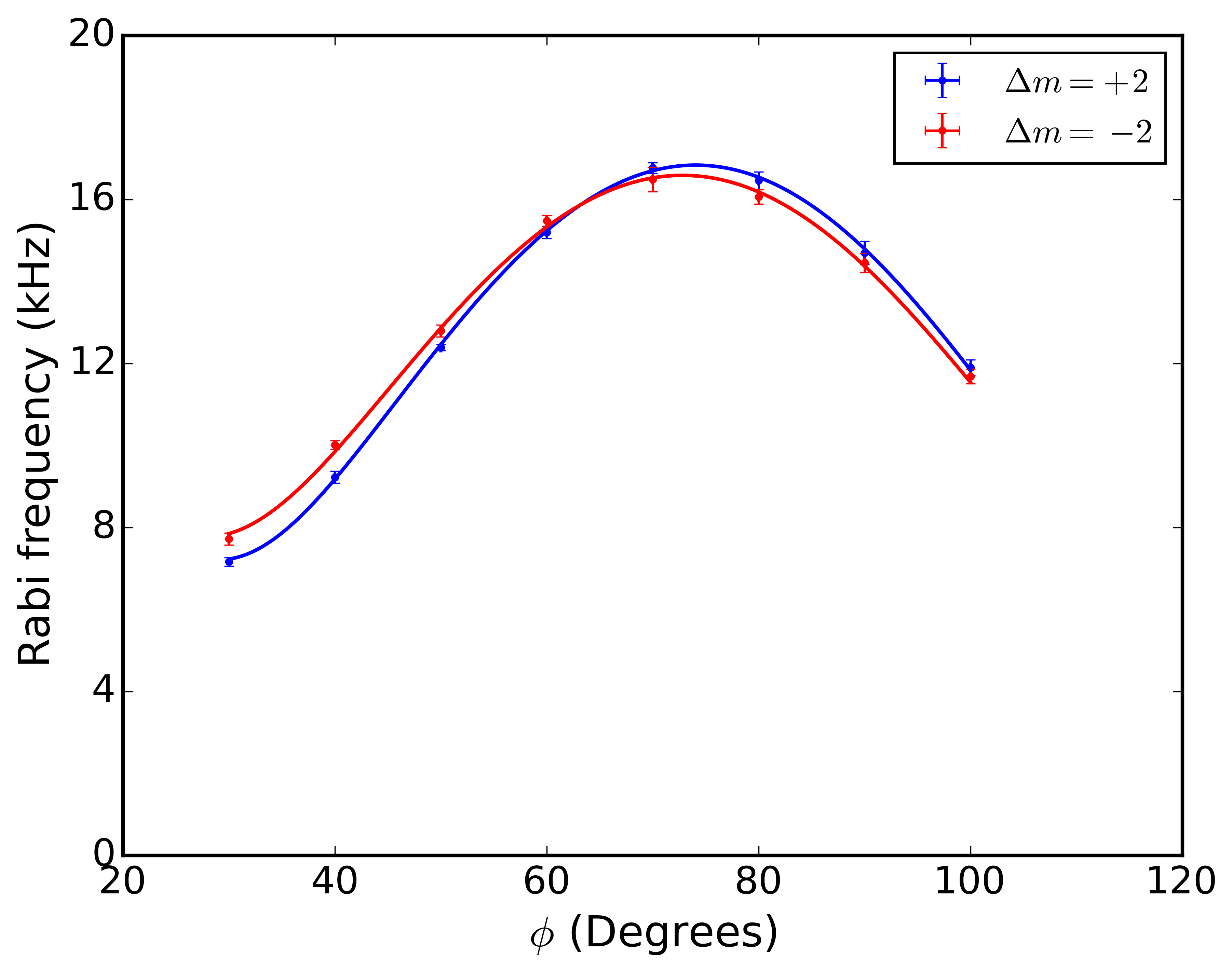}
  \caption{The data and the fit of the $\Omega^{(\pm 2)}$ measurements fitted simultaneously to the birefringence laser field model.  The blue indicates the $\Delta m$ = +2 data and the red indicates the $\Delta m$ = -2 data.}    
  \label{Fig:PM2AndSimulfit}
\end{figure}

%
Stress in optical windows is well known to induce birefringence which can hinder precision measurements.  
Several recent papers have addressed the topic with methods for mitigating or measuring the effect \cite{Brakhane15, Solmeyer11, Steffen13}, however for our purposes a new technique was required.
When driven with an elliptically polarized field, the minimum in the $\Delta$m = 0 transition, $\Omega^{(0)}_{min}$, acquires an additional contribution from electric-quadrupole coupling, $\Omega^{(\epsilon)}_{min}$, which from Eq. (\ref{eq:Rabiform}) is, 
\begin{equation}\label{Eq:ToGetEllipCorrect}
	\begin{split}
		\Omega^{(0)}_{min} = \Big| \frac{ik}{4\sqrt{10}\hbar}\, &E2 \, \sin(2\theta)E_{||}(|E|,\,\phi_{0},\,\alpha,\,\Gamma) \\
		\,- \frac{1}{\sqrt{6}\hbar}\, &\mathrm{M1}\, \sin(\theta) B_{||}(|E|,\,\phi_{0},\,\alpha,\,\Gamma) \Big|		
	\end{split}
\end{equation}
\begin{equation*}
	= \Big| \Omega^{(\epsilon)}_{min} + A^{(\mathrm{M1})}_{min}\times \mathrm{M1} \Big|
\end{equation*}
and from which M1 can be calculated once the viewport optical axis orientation, $\alpha$, and its retardance, $\Gamma$ are known. \\
\indent To experimentally determine the viewport optical parameters, we measured $\Omega^{(\pm 2)}(\phi)$ concurrently.  For a perfectly linear polarized beam these transition's Rabi frequencies are identical, however, when driven with an elliptically polarized field, interference between the different couplings causes the $\Omega^{\pm 2}$($\phi$) curves to become skewed with respect to one another.  Where the two curves cross over each other indicates the orientation of the optical axes, and their relative skew increases with $\Gamma$.  The data are plotted in Fig. (\ref{Fig:PM2AndSimulfit}) where the blue and red curves there are $\Delta m$ of +2 and -2, respectively. \\         
\indent These data were collected and interpreted in isolation from the previous data sets because of changes that were required of the system for another experiment.  However, the 2.051 $\mu$m pointing through the viewport was not changed.  The data were collected by fixing the beam's polarization and measuring both of the $\pm$ 2 transition Rabi frequencies before rotating the beam's polarization. \\
\indent The precision to which $\alpha$ and $\Gamma$ are extracted can be greatly improved by taking combinations of the data sets that exaggerate the effect of those parameters by masking the influence of other extraneous parameters.  The first of these combinations is the ratio $\Omega^{(+2)}/ \Omega^{(-2)}$, which for perfect linear polarized light does not deviate from unity.  Because in taking the ratio the electric field amplitude is canceled, to first order any deviation of $\Omega^{(+2)}/ \Omega^{(-2)}$ from unity is due to $\Gamma$ alone. The purpose of measuring both $\pm$ 2 Rabi frequencies at a given half-wave plate orientation before moving to the next position was to minimize the possibility that the laser field amplitude could have changed significantly between the scans.  To that concern though, over the several years during which this measurement was refined, there was never any indication that such laser field changed meaningfully over the course of an experiment.  The second combination of the data is the squared difference 
between the two data sets, $\big|\Omega^{(+2)}\big|^{2} \,-\, \big|\Omega^{(-2)}\big|^{2}$, which simplifies to,
\begin{equation}\label{Eq:SquareDiff}
  \big|\Omega^{(+2)}\big|^{2} \,-\, \big|\Omega^{(-2)}\big|^{2} = 2\,A\,\sin\big[4(\phi\,-\,\phi_{0}) - 2\alpha \big] 
\end{equation}
The coefficient $A=k/(4\sqrt{10}\hbar)|E|E2\,\sin(2\theta)\cos(\theta)\sin(\Gamma)$, and does not provide information about those parameters, but the combination is useful for obtaining $\alpha$.  To do this we use the known value of $\phi_{0}$, which leaves $\alpha$ as a second free parameter of the model.  From these two combinations of the data we obtain the viewport parameters that are given in the table in Fig. (\ref{Fig:PM2AndSimulfit}), and additionally with the parameters in Table (\ref{Table:Rabis0P2FitParameters}) we find, 
\begin{figure}
  \centering
  \includegraphics[scale=0.35]{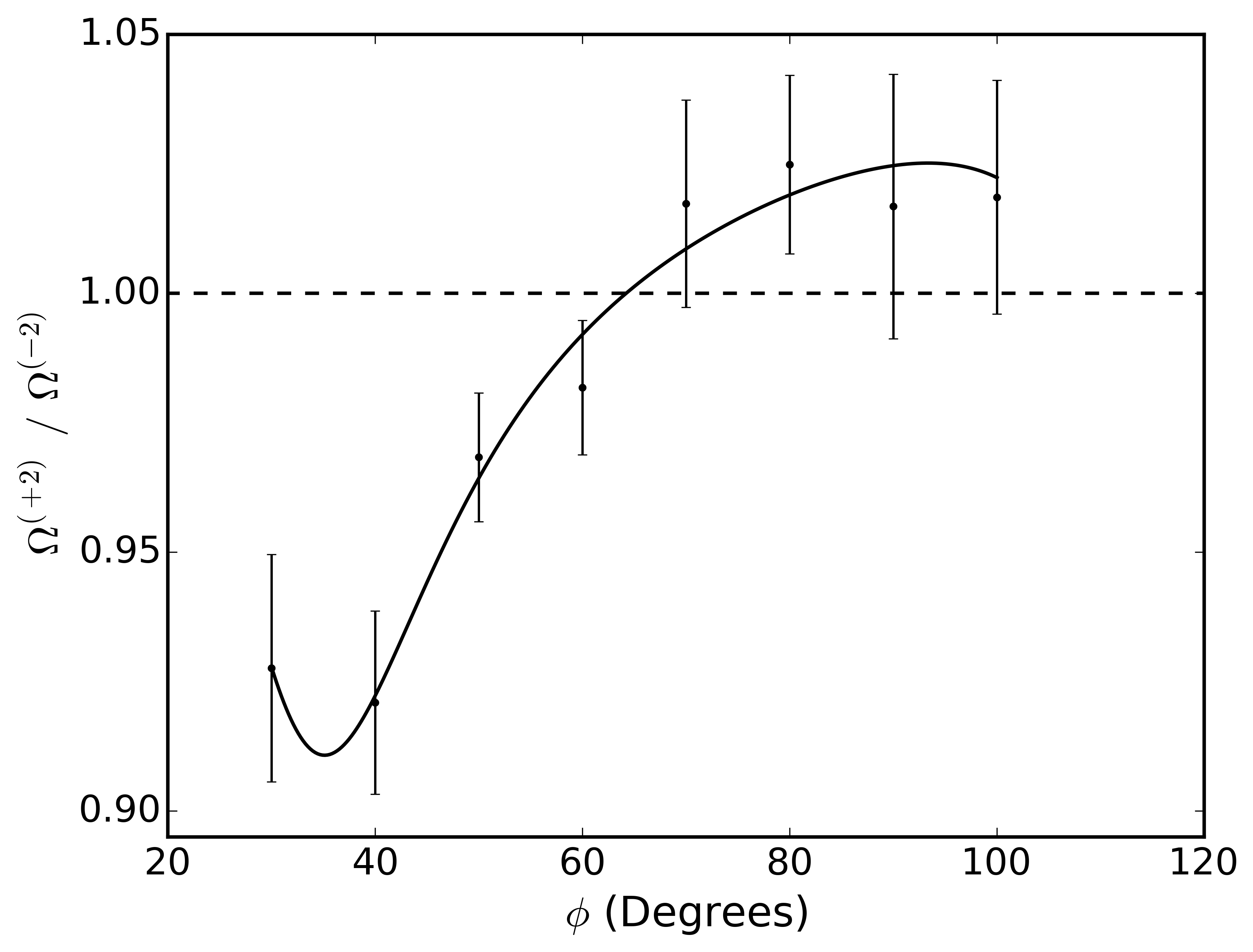}
  \includegraphics[scale=0.35]{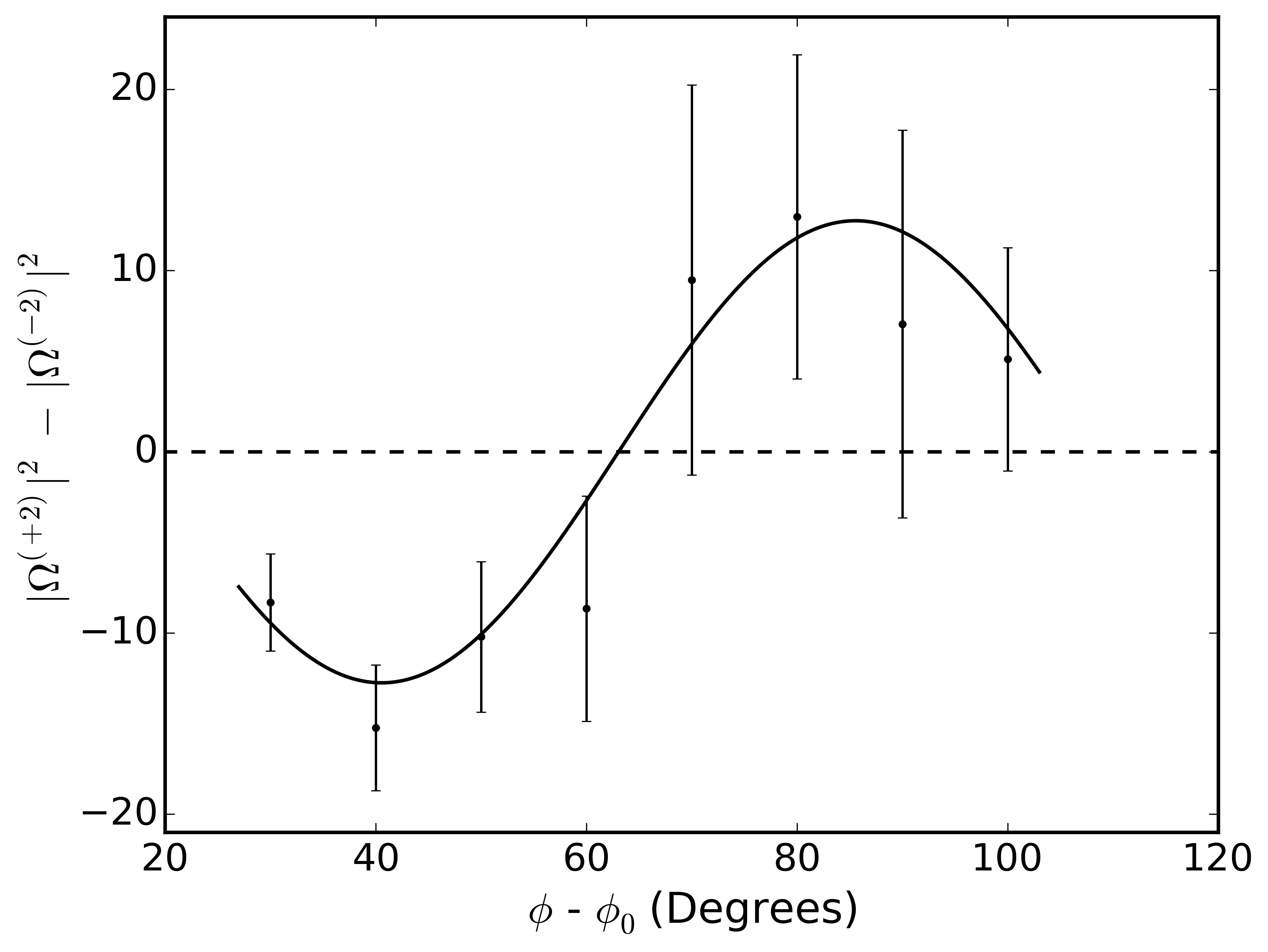}\\
  \quad \\
  \begin{tabular}{c c c c c} 
	  \hline\hline\vspace{2 mm} 
	  $\Omega^{(+2)}/ \Omega^{(-2)}$ & $\Gamma$ (Degrees) & $\alpha$ (Degrees) \\
	  \hline
	  & $2.8^{\circ} \pm 0.3^{\circ}$ & $155^{\circ} \pm 9^{\circ}$ \\
	  & & \\
	  \hline\hline 
	  \vspace{2 mm}
	  $\big|\Omega^{(+2)}\big|^{2} \,-\, \big|\Omega^{(-2)}\big|^{2}$ & $A$ (Hz$^{2}$) & $\alpha$ (Degrees) \\
	  \hline 
	  & $3.57 \pm .07$ & $159^{\circ} \pm 1.6^{\circ}$ \\
	  \hline\hline 
  \end{tabular}
\caption{The top panel shows a plot of the ratio between the $\Omega^{(+ 2)}$ and $\Omega^{(- 2)}$ data sets and the bottom panel shows a plot of the square difference $\big|\Omega^{(+2)}\big|^{2} \,-\, \big|\Omega^{(-2)}\big|^{2}$.  From these curves we are able to extract the viewport optical axis orientation and retardance experienced by the 2.051 $\mu$m beam, which are listed in the table.  }
\label{Fig:RatioandSquareDiffDataAndFits}
\end{figure}
\begin{equation}
  \begin{split}
    \mathrm{M1}& < 93 \pm 39 \times 10^{-5} \mu_{B} \\
    (\text{with } &\text{ellipticity correction})
  \end{split}
\end{equation}  
\indent Our reported uncertainty is from the quadrature sum of those from $\Omega^{(0)}_{min}$, $\Omega^{(\epsilon)}_{min}$, and $A^{(\mathrm{M1})}_{min}$.
Table (\ref{Table:FinalPropagate}) summarizes the error propagation.  The uncertainty from measuring $\Omega^{(0)}_{min}$ is the largest single contributor, with 
that from $\Omega^{(\epsilon)}_{min}$ contributing an almost equal portion.  In the latter, as well as in $A^{(\mathrm{M1})}_{min}$, the uncertainty in the calculated
fields dominate, which are themselves dominated by the uncertainty in $\Gamma$. \\
\begin{table}
  \caption{A summary of the propagated uncertainty to M1.  The parameters defined in Eq. (\ref{Eq:ToGetEllipCorrect}) are demarked by $Y$.  The largest fractional contributors
  to their uncertainty are given in the $X$ column.  The relative uncertainty in $E2$ is estimated from \cite{Sahoo06} to be 1\%.  The uncertainty in $\big|E_{||}\big|$ and $\big|B_{||}\big|$ were obtained 
  by propagating the uncertainties from $\big|E\big|$, $\phi_{0}$, $\alpha$, and $\Gamma$.  } 
  \label{Table:FinalPropagate}
  \bgroup
  \setlength\tabcolsep{7pt}
  \centering 
  \begin{tabular}{c c c c c} 
    \hline\hline\vspace{2 mm} 
    $Y/2\pi$ & $X$ & \big|$\partial Y$/$\partial X \big|$ $\sigma_{X}$/$Y$ & $\big| \partial \mathrm{M1}$/$\partial Y \big|$ $\sigma_{Y}$  \\ [0.5ex]
    \hline
    $\Omega^{(0)}_{min}$ & $--$ & $--$  & 28 $\times 10^{-5} \mu_{B}$\\
    \hline
    $\Omega^{(\epsilon)}_{min}$ & & \\
				&  $|E_{||}|$ & 0.148 & \\			
				&  $E2$      & 0.010 & \\
				&  $\theta$  & 0.005 & \\   			                            
				& &          & 24 $\times 10^{-5} \mu_{B}$ \\
    \hline 
    $A^{(\mathrm{M1})}_{min} $ &  &  &  \\
		      &  $|B_{||}|$ & 0.029 &\\			
		      &  $\theta$  & 0.003 &\\
		      & &          & 13 $\times 10^{-5} \mu_{B}$ \\
\hline
  & & Total $\sigma_{\mathrm{M1}}$ & 39 $\times 10^{-5} \mu_{B}$\\
    \hline\hline  
  \end{tabular} 
  \egroup
\end{table}	  
\section{Conclusion}
\indent By performing a polarization-based spectroscopy of the 6S$_{1/2} \leftrightarrow$ 5D$_{3/2}$ transitions in Ba$^{+}$ we have placed an upper bound
on those transition's magnetic-dipole moment.  In the course of doing so we have also demonstrated how our technique can be used to measure stress-induced birefringence
in ultra-high vacuum viewports.  Several pathways for improvement exist for a next-generation of this experiment.  
To improve the ability to resolve $\Omega^{(0)}_{min}$ efforts are underway to reduce our measurement's decoherence rate, which was in majority set by ambient magnetic field noise.  This could be mitigated with the implementation of magnetic shielding.  A complementary strategy would be to enlarge $\Omega^{(0)}_{min}$ by delivering more 2$\mu$m light to the ion.
Also, as our measurement has shown, ellipticity can be leveraged - if sufficiently well calibrated - to enhance the size of $\Omega^{(0)}_{min}$.  This approach could be improved by eliminating stress-birefringence from the system and introducing a large but known amount of ellipticity separately.  This reasoning leads naturally to a version of our measurement using 
a circularly polarized 2 $\mu$m laser field, which we have proposed in \cite{Williams13} and \cite{Williams15}.\\
\indent The authors thank the other members of the University of Washington's Trapped Ion Quantum Computing group for 
their varied and valuable assistance throughout, and also the National Science Foundation for their generous support during the early stages of this work. \\

\end{document}